\documentclass[twocolumn]{aastex701}

\usepackage{amsmath}

\newcommand{\Reff}{R_{\mathrm{eff}}}
\newcommand{\SSFR}{\Sigma_{\mathrm{SFR}}}
\newcommand{\Msun}{M_{\odot}}
\newcommand{\Ha}{\mathrm{H}\alpha}

\shorttitle{Local $\Sigma_{\rm SFR}$ and Young Cluster Radii}
\shortauthors{Sebastian}

\begin{document}

\title{The Local Star Formation Environment Predicts Young Star Cluster Radii:
Evidence from PHANGS--MUSE H$\alpha$ Maps}

\author{Michelle Sebastian}
\email{michelle.m.sebastian@gmail.com}
\affiliation{Simi Valley High School, Simi Valley, CA 93065, USA}

\begin{abstract}

The effective radii of young star clusters correlate with cluster mass, but the
large scatter about that relation is not well understood. Theory predicts that a
cluster's initial radius depends on the gas surface density of its natal
environment, so the star formation environment should leave a measurable imprint
on cluster structure. In a companion analysis of 5,105 clusters across 31 LEGUS
galaxies, we found that galaxy-wide specific star formation rate (sSFR) carries
no predictive power for cluster radius. Here we test whether that null result
reflects the spatial scale on which the environment was measured. We cross-match
the LEGUS cluster catalog of Brown \& Gnedin with H$\alpha$ maps from the
PHANGS--MUSE survey in the four galaxies common to both surveys, and measure the
local star formation rate surface density, $\Sigma_{\rm SFR}$, at the position
of every cluster by aperture photometry. For the 325 clusters younger than
10~Myr with reliable radius and mass measurements, local $\Sigma_{\rm SFR}$
significantly predicts effective radius ($\beta = 0.079$, $p = 2\times10^{-4}$)
in a model containing mass and age, while the host galaxy's sSFR does not
($\beta = 0.042$, $p = 0.40$) for those same clusters. The local model
generalizes nearly three times better under cross-validation
(CV~$R^2 = 0.061$ versus $0.022$). The effect operates among clusters within a
single galaxy: it survives galaxy fixed effects ($\beta = 0.068$, $p = 0.001$)
and appears independently in each of the two galaxies large enough to fit alone.
Including local $\Sigma_{\rm SFR}$ also reduces the mass coefficient by 26\%.
The birth environment measurably shapes cluster structure, but only when it is
measured on the scale on which clusters actually form.

\end{abstract}


\section{Introduction} \label{sec:intro}

Star clusters are among the most useful tracers of how galaxies assemble their
stellar mass. Because the stars in a cluster share an age, a metallicity, and a
formation site, the structural properties of clusters record the conditions
under which star formation proceeded \citep{PortegiesZwart2010, Krumholz2019}.
The scaling between cluster effective radius, $\Reff$, and cluster mass is the
most widely studied of these structural relations, and what sets it, and what
sets its scatter, remains an open question.

The simplest expectation follows from virial equilibrium. For a self-gravitating
system in which $2K + U = 0$, mass, radius, and velocity dispersion are linked by
$M \propto \sigma^2 R$, so that mass is the primary structural parameter. Observed
relations are considerably shallower than this argument alone would suggest.
\citet{BrownGnedin2021} fit \citet{Elson1987} surface brightness profiles to
6,097 clusters in the 31 galaxies of the Legacy ExtraGalactic UV Survey
\citep[LEGUS;][]{Calzetti2015}, building on earlier size measurements by
\citet{Ryon2017}, and recovered $\Reff \propto M^{0.24}$ with an intrinsic
scatter of $0.25$~dex. That scatter is large: at fixed mass, cluster radii span
close to an order of magnitude. Something other than mass is clearly at work.

Theory points toward the formation environment. \citet{ChoksiKruijssen2021}
predict that the initial half-mass radius of a cluster should scale as
$r_h \propto m^{1/2}\,\Sigma_{\rm gas}^{-1/2}$, so that clusters forming in
denser gas are born more compact. Related work on the cluster formation
efficiency reaches a similar conclusion, namely that the surface density and
pressure of the natal interstellar medium govern how star formation is organized
into bound systems \citep{Kruijssen2012, Chevance2020}. If these predictions
hold, some measure of the star-forming environment ought to carry predictive
power for cluster radius beyond mass alone.

In \citet[][hereafter Paper~I]{Sebastian2026} we tested exactly this, using the
galaxy-wide specific star formation rate (sSFR) of the host galaxy as the
environmental variable. Across 5,105 LEGUS clusters with reliable radius and
mass measurements, sSFR contributed essentially nothing
($\beta = 0.0002 \pm 0.015$, $p = 0.99$), and a multivariate model including
sSFR and age generalized \emph{worse} than a mass-only model under
cross-validation (CV~$R^2 = -0.017$ versus $0.028$). Taken at face value, that
result suggests the star formation environment is irrelevant to cluster
structure, in direct tension with \citet{ChoksiKruijssen2021}.

There is an alternative reading. A galaxy-wide sSFR is a single number averaged
over tens of kiloparsecs. Clusters form inside individual giant molecular
clouds, on scales of tens of parsecs, and the conditions in those clouds vary
enormously within one galaxy \citep{Sun2018}. Every cluster in a given LEGUS
galaxy is assigned the same sSFR by construction, so any environmental
dependence that operates below the scale of the galaxy is averaged away before
the regression ever sees it. The null result of Paper~I may therefore be a
statement about measurement scale rather than about physics.

This paper tests that interpretation directly. We measure the star formation
rate surface density, $\SSFR$, at the position of each individual cluster from
$\Ha$ emission maps taken by the PHANGS--MUSE survey \citep{Emsellem2022}, and
ask whether this local quantity predicts cluster radius where the global one did
not. The comparison is deliberately constructed to be as close to
like-for-like as possible: both environmental measures are tested on the same
clusters, in the same regression framework, with the same covariates. To our
knowledge this is the first direct head-to-head test of local against global
star formation environment as a predictor of individual cluster radius.

Section~\ref{sec:data} describes the two surveys and the construction of the
sample. Section~\ref{sec:methods} sets out the photometry, the conversion to
$\SSFR$, and the regression framework. Section~\ref{sec:results} presents the
head-to-head comparison and robustness tests, Section~\ref{sec:discussion}
interprets the result and states its limitations, and
Section~\ref{sec:summary} summarizes.

\section{Data} \label{sec:data}

\subsection{The LEGUS Cluster Catalog} \label{sec:legus}

Cluster properties are taken from the catalog of \citet{BrownGnedin2021}, which
provides homogeneous radii for 6,097 clusters in the 31 LEGUS galaxies
\citep{Calzetti2015}. Radii are measured by fitting \citet{Elson1987} profiles,
generalized to include ellipticity, to \textit{Hubble Space Telescope} imaging,
with uncertainties from bootstrap resampling. Cluster masses and ages come from
the LEGUS spectral energy distribution fits \citep{Adamo2017, Cook2019}. Galaxy
distances follow the tip-of-the-red-giant-branch measurements of
\citet{Sabbi2018}, and host galaxy stellar masses, star formation rates, and
specific star formation rates are from \citet{Calzetti2015}. Only morphological
class 1 and 2 objects, that is compact and centrally concentrated sources, are
included in the catalog.

Two quality flags accompany each entry. The \texttt{reliable\_radius} flag
identifies clusters whose profile fit did not fail and which are not in the
worst tenth percentile of the fit quality metric; \texttt{reliable\_mass}
identifies clusters whose spectral energy distribution fit passes a $Q$-statistic
check. \citet{BrownGnedin2021} use only clusters satisfying both flags, and so
do we.

\subsection{PHANGS--MUSE H$\alpha$ Maps} \label{sec:phangs}

Local star formation rates are measured from the PHANGS--MUSE survey
\citep{Emsellem2022}, which provides optical integral field spectroscopy of
nearby star-forming galaxies obtained with the Multi Unit Spectroscopic Explorer
\citep[MUSE;][]{Bacon2010} on the Very Large Telescope. We use the convolved
optimized (\texttt{copt}) MAPS data products, in which each galaxy mosaic has been
convolved to a single, spatially uniform point spread function across that
mosaic, and extract the
\texttt{HA6562\_FLUX} and \texttt{HA6562\_FLUX\_ERR} extensions. The data were
retrieved from the CANFAR science portal.

$\Ha$ recombination emission at $6562.8$~\AA{} arises from hydrogen ionized by
massive, short-lived stars. It therefore traces star formation over
approximately the past 10~Myr \citep{KennicuttEvans2012}, which is precisely the timescale relevant to the
clusters considered here.

\subsection{Cross-Matching and Sample Construction} \label{sec:sample}

\begin{figure*}
\centering
\includegraphics[width=\textwidth]{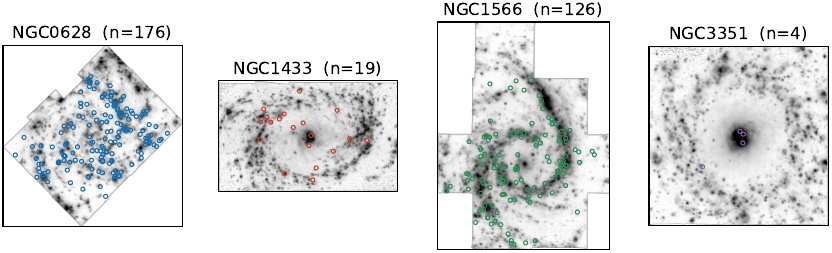}
\caption{PHANGS--MUSE H$\alpha$ flux maps of the four galaxies covered by both
LEGUS and PHANGS--MUSE, with the positions of the young ($<10$~Myr) LEGUS
clusters in the final reliable sample overlaid. Clusters trace the bright
\ion{H}{2} complexes in the spiral arms and circumnuclear rings, which is the
variation in local star formation environment that a single galaxy-wide sSFR
value averages away. The grayscale is logarithmic in H$\alpha$ flux, with dark
denoting bright emission; marker color encodes the host galaxy and is used
consistently in Figures~\ref{fig:dynrange} and \ref{fig:partial}. Panel titles
give the number of clusters in the final sample (Table~\ref{tab:galaxies}).
\label{fig:maps}}
\end{figure*}

Four galaxies are covered by both surveys: NGC~628, NGC~1433, NGC~1566, and
NGC~3351. Their properties are summarized in Table~\ref{tab:galaxies} and their
H$\alpha$ maps are shown in Figure~\ref{fig:maps}.

LEGUS cluster identifiers are assigned field by field and are not unique across
the survey, so clusters were matched to the $\Ha$ maps by sky position
(right ascension and declination) rather than by identifier, and the
\citet{BrownGnedin2021} reliability flags were propagated through the same
positional match.

The sample was then built in three steps.

\begin{enumerate}
\item \emph{Spatial overlap.} Clusters in the four galaxies above were retained.
\item \emph{Age cut.} Because $\Ha$ traces only the past ${\sim}10$~Myr, clusters
  older than this may have drifted from their birth sites, in which case the
  $\Ha$ flux at their present position is not a proxy for the conditions in
  which they formed. Requiring an age below 10~Myr, and a finite positive
  $\Ha$ flux in the aperture, leaves 618 clusters.
\item \emph{Reliability cut.} Requiring both \texttt{reliable\_radius} and
  \texttt{reliable\_mass} reduces the sample to 325 clusters.
\end{enumerate}

The final sample is strongly unbalanced between hosts. NGC~628 (176 clusters)
and NGC~1566 (126) together account for 93\% of it, while NGC~1433 contributes
19 clusters and NGC~3351 four. Any statement in this paper about ``four
galaxies'' should be read with that in mind, and Section~\ref{sec:within}
reports the tests that this imbalance makes necessary.

We adopt the 325-cluster sample for all primary results, because it matches the
measurement quality of the Paper~I baseline and is the more conservative choice.
The 618-cluster sample is retained as a robustness check
(Section~\ref{sec:robustness}). The per-galaxy breakdown of both samples is
given in Table~\ref{tab:galaxies}.

\begin{deluxetable*}{lccccc}
\tablecaption{Galaxies with Both LEGUS and PHANGS--MUSE Coverage
\label{tab:galaxies}}
\tablehead{
\colhead{Galaxy} & \colhead{Distance} & \colhead{Galaxy sSFR} &
\colhead{$N$ young} & \colhead{$N$ reliable} & \colhead{1\arcsec{} scale} \\
\colhead{} & \colhead{(Mpc)} & \colhead{(Gyr$^{-1}$)} &
\colhead{($<10$~Myr)} & \colhead{(final)} & \colhead{(pc)}
}
\startdata
NGC 628  & 8.80  & 0.334 & 258 & 176 & 42.7 \\
NGC 1433 & 9.10  & 0.016 &  68 &  19 & 44.1 \\
NGC 1566 & 15.60 & 0.210 & 279 & 126 & 75.6 \\
NGC 3351 & 9.30  & 0.075 &  13 &   4 & 45.1 \\
\hline
Total    & \nodata & \nodata & 618 & 325 & \nodata \\
\enddata
\tablecomments{Distances and galaxy-wide specific star formation rates are as
adopted by \citet{BrownGnedin2021}, following \citet{Sabbi2018} and
\citet{Calzetti2015}. The final column gives the physical radius subtended by
the 1\arcsec{} photometric aperture at the distance of each galaxy. Note the
strong imbalance: NGC~628 and NGC~1566 together supply 302 of the 325 clusters
in the final sample, while NGC~3351 supplies four.}
\end{deluxetable*}

\section{Methods} \label{sec:methods}

\subsection{Local Star Formation Rate Surface Density} \label{sec:sigmasfr}

For each cluster we performed circular aperture photometry on the $\Ha$ flux map
at the cluster position using \texttt{photutils} \citep{photutils}, with an
aperture radius of $1\arcsec$. This radius is a compromise: it is large enough to
average over the spaxel-to-spaxel noise of the MUSE mosaics and over small
astrometric offsets between the \textit{HST} and MUSE frames, and small enough
to remain local to the cluster. At the distances of these four galaxies it
corresponds to physical radii of 42.7 to 75.6~pc (Table~\ref{tab:galaxies}).

The aperture flux $F_{\Ha}$ was converted to a luminosity using the adopted
galaxy distance $D$,
\begin{equation}
L_{\Ha} = 4\pi D^{2} F_{\Ha},
\end{equation}
and then to a star formation rate using an $\Ha$ calibration of the
\citet{Kennicutt1998} form,
\begin{equation}
\mathrm{SFR}\,[\Msun\,\mathrm{yr^{-1}}] = C\,
L_{\Ha}\,[\mathrm{erg\,s^{-1}}],
\end{equation}
with $C = 5.5\times10^{-42}\,\Msun\,\mathrm{yr^{-1}\,(erg\,s^{-1})^{-1}}$.
The value of $C$ depends on the assumed stellar initial mass function (IMF),
because $\Ha$ traces only the ionizing output of the most massive stars and $C$
encodes how much total stellar mass accompanies them.
\citet{Kennicutt1998} derived $C = 7.9\times10^{-42}$ for a
Salpeter IMF; the value adopted here corresponds instead to the Kroupa-type
IMFs in standard current use, which contain fewer low-mass stars per ionizing
star and therefore require a smaller constant by a factor of about 1.6
\citep{KennicuttEvans2012}. The choice of IMF rescales every $\SSFR$ in the
sample by the same factor and so, as noted above, affects only the fitted
intercept.
Dividing by the physical area subtended by the aperture gives the local star
formation rate surface density,
\begin{equation}
\SSFR = \frac{\mathrm{SFR}}{\pi R_{\mathrm{ap}}^{2}},
\end{equation}
in $\Msun\,\mathrm{yr^{-1}\,pc^{-2}}$, where $R_{\rm ap}$ is the aperture radius
in parsecs. Because $\SSFR$ enters the regression only in logarithmic form, any
multiplicative constant common to all clusters shifts the fitted intercept but
leaves the $\SSFR$ coefficient, its standard error, and its $p$-value unchanged.
The galaxy-dependent factor $D^2$ does not cancel in this way, and is applied
per galaxy.

No correction for internal dust extinction was applied; the consequences are
discussed in Section~\ref{sec:limitations}.

\subsection{Regression Framework} \label{sec:regression}

All quantities were transformed to base-10 logarithms so that power-law
relations become linear, and models were fitted by ordinary least squares. The
general form is
\begin{equation}
\log_{10}\Reff = \beta_0 + \beta_1 \log_{10} M + \beta_2 \log_{10}\tau
+ \beta_3 \log_{10} E + \varepsilon,
\label{eq:model}
\end{equation}
where $M$ is cluster mass, $\tau$ is cluster age, and $E$ is the environmental
term. The entire comparison in this paper turns on a single substitution in
Equation~\ref{eq:model}:
\begin{description}
\item[Model A (global)] $E$ is the specific star formation rate of the host
  galaxy, identical for every cluster in that galaxy.
\item[Model B (local)] $E$ is $\SSFR$ measured in the aperture centered on that
  individual cluster.
\end{description}
Both models are fitted to the same 325 clusters and contain the same mass and
age terms, so any difference between them is attributable to the environmental
variable alone.

\subsection{Model Comparison} \label{sec:comparison}

Following Paper~I, models are assessed by 5-fold cross-validation rather than by
in-sample fit statistics. The data are split into five folds; each fold in turn
is held out while the model is fitted to the remaining 80\% and scored on the
held-out 20\%. The reported CV~$R^2$ is the mean across folds. Folds were shuffled
with a fixed random seed so that the comparison between models uses identical
splits. Training $R^2$ is reported alongside, but the gap between the two is the
diagnostic of interest: a model that captures genuine signal loses little
predictive power out of sample, whereas a model fitting noise loses much.

The statistical significance of each environmental term was assessed with a
partial $F$-test comparing the three-predictor model against the mass-and-age
model nested within it. Variance inflation factors were computed to confirm that
the predictors are not strongly collinear. Analyses used
\texttt{scikit-learn} \citep{sklearn}, \texttt{statsmodels} \citep{statsmodels},
\texttt{astropy} \citep{Astropy2022}, \texttt{numpy} \citep{numpy},
\texttt{scipy} \citep{scipy}, and \texttt{matplotlib} \citep{matplotlib}.

\section{Results} \label{sec:results}

\subsection{Local Versus Global Environment} \label{sec:headtohead}

Table~\ref{tab:coeffs} gives the fitted coefficients of Models A and B for the
325 reliable young clusters, and Table~\ref{tab:performance} gives the
corresponding model-level statistics.

\begin{figure}
\centering
\includegraphics[width=\columnwidth]{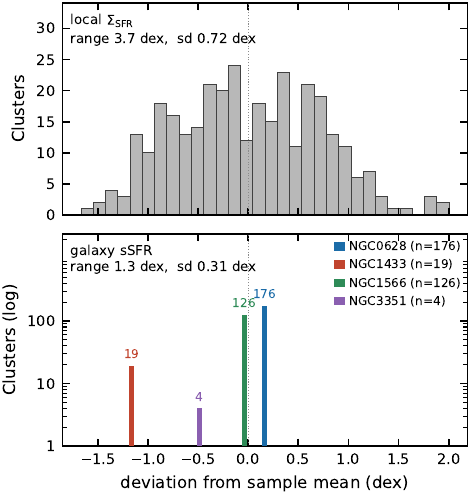}
\caption{The two environmental variables measured on the same 325 clusters,
each plotted as a deviation from its own sample mean. The quantities carry
different physical units ($\Msun\,\mathrm{yr^{-1}\,pc^{-2}}$ and
$\mathrm{yr^{-1}}$) and so cannot be compared in absolute terms; centering each
on its own mean isolates the property the regression actually responds to,
namely how much it varies; the standard deviation of each is annotated on its
panel. \textit{Top:} local $\Sigma_{\rm SFR}$, spanning 3.7~dex, on a linear
count axis. \textit{Bottom:} galaxy-wide sSFR, which takes only four values, one
per host. The count axis here is logarithmic, without which NGC~3351 (four
clusters) would be invisible beside NGC~628 (176); the number of clusters is
printed above each bar. The two galaxies supplying 93\% of the sample differ by
just 0.20~dex. A variable that barely varies within the sample cannot predict
anything, which is the reason for the contrast in
Section~\ref{sec:headtohead}.
\label{fig:dynrange}}
\end{figure}

The contrast is clear. The galaxy-wide sSFR coefficient is $0.042$ with
$p = 0.40$: on this sample there is no evidence that the overall star formation
activity of a galaxy predicts the sizes of the individual clusters within it,
which reproduces the Paper~I null result. These 325 clusters are a subset of
the 5,105 analyzed in Paper~I rather than an independent sample, so this is a
consistency check rather than a confirmation, but it establishes that the global
variable fails on exactly the clusters where the local one succeeds.
Figure~\ref{fig:dynrange} makes the underlying reason plain: across these 325
clusters the local measurement spans 3.7 dex, from $\log_{10}\SSFR = -9.78$ to
$-6.11$ in $\Msun\,\mathrm{yr^{-1}\,pc^{-2}}$, while the global variable takes
exactly four values, one per host galaxy. The disparity is sharper than those
figures suggest: NGC~628 and NGC~1566, which between them supply 93\% of the
sample, differ by only 0.20~dex in galaxy-wide sSFR. For almost every cluster
in the sample, the global variable is very nearly a constant.
Replacing that term with the local $\SSFR$ measured at each cluster's own
position yields a coefficient of $0.079$ with $p = 2\times10^{-4}$. The local
coefficient is $1.9$ times the global one and is significant at high confidence,
while the global one is consistent with zero.

\begin{figure}
\centering
\includegraphics[width=\columnwidth]{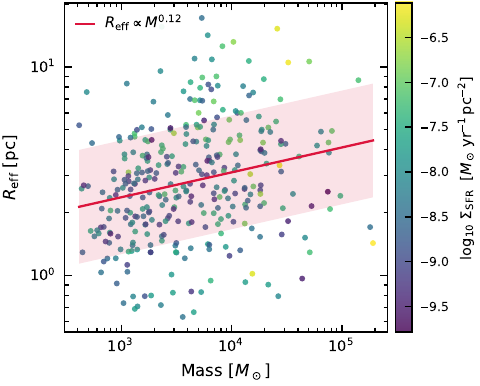}
\caption{Effective radius against mass for the 325 reliable young clusters,
colored by local star formation rate surface density. The solid line is the
mass-only fit and the shaded band its $\pm1\sigma$ scatter. At fixed mass,
clusters in regions of higher local $\Sigma_{\rm SFR}$ sit preferentially above
the relation.
\label{fig:massradius}}
\end{figure}

\begin{figure}
\centering
\includegraphics[width=\columnwidth]{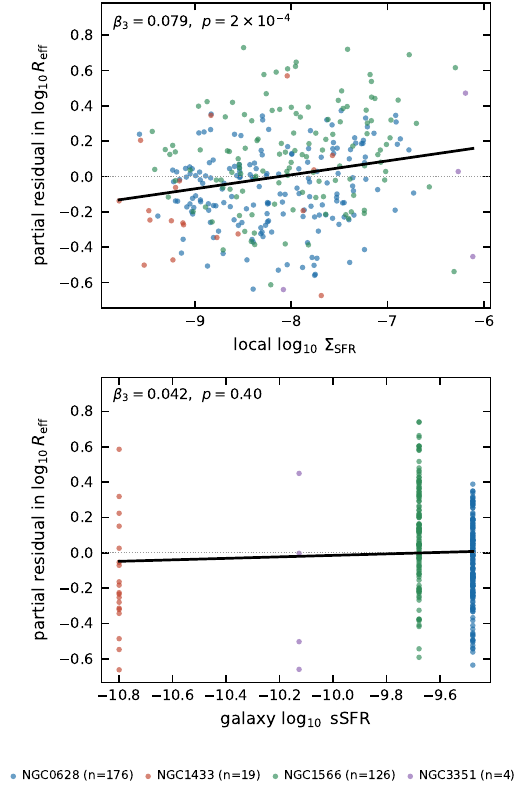}
\caption{Partial residual plot for the environmental term. Each panel shows
$e_i + \beta_3 x_i$ against $x_i$, where $e_i$ are the residuals of the full
three-predictor model; constructed this way the fitted slope equals the
tabulated coefficient exactly. \textit{Top:} local
$\log_{10}\Sigma_{\rm SFR}$, slope $\beta_3 = 0.079$ ($p = 2\times10^{-4}$).
\textit{Bottom:} galaxy-wide $\log_{10}$ sSFR in $\mathrm{yr^{-1}}$
(Table~\ref{tab:galaxies} tabulates the same quantity in
$\mathrm{Gyr^{-1}}$), slope $\beta_3 = 0.042$ ($p = 0.40$), where the four
discrete host values collapse the sample into vertical stripes. Points are colored by host galaxy. This is the head-to-head
comparison of Section~\ref{sec:headtohead} in a single figure.
\label{fig:partial}}
\end{figure}

Figure~\ref{fig:massradius} shows the mass-radius plane for this sample colored
by local $\SSFR$. A mass-only fit to these 325 clusters gives
$\Reff \propto M^{0.12}$, shallower than the $M^{0.24}$ that
\citet{BrownGnedin2021} recover for the catalog as a whole. This is expected
rather than discrepant: they report a shallower slope for clusters younger than
10~Myr, and every cluster here is younger than 10~Myr by construction.
Figure~\ref{fig:partial} isolates the environmental effect directly. Once the
contributions of mass and age are accounted for, the partial residual in
$\log_{10}\Reff$ still rises systematically with $\log_{10}\SSFR$: clusters
forming in more intensely star-forming regions are larger than mass and age
alone would predict. The equivalent panel for galaxy sSFR shows no significant
trend and, because every cluster in a galaxy is assigned the same value,
collapses into four vertical stripes.

\begin{figure}
\centering
\includegraphics[width=\columnwidth]{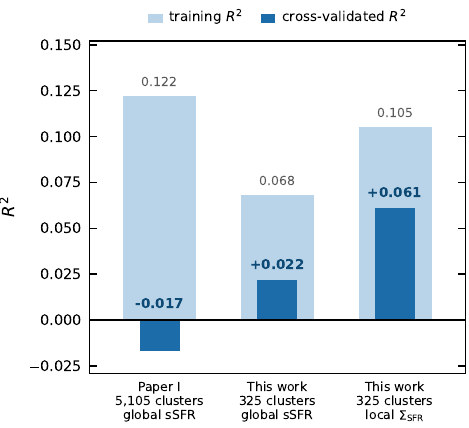}
\caption{Training $R^2$ (faded) and cross-validated $R^2$ (solid) for three
analyses. \textit{Left:} the Paper~I full model on 5,105 clusters with
galaxy-wide sSFR, which overfits (CV~$R^2 = -0.017$). \textit{Center:} the same
model form restricted to the 325 reliable young clusters, still with global sSFR
(CV~$R^2 = 0.022$). \textit{Right:} the identical sample with global sSFR
replaced by local $\Sigma_{\rm SFR}$ (CV~$R^2 = 0.061$). Only the center-to-right
step is a controlled comparison: those two models use the identical 325 clusters
and differ solely in the environmental variable. The left-to-center step changes
the age cut, the number of host galaxies and the sample size at once, so no part
of it can be attributed to any single change.
\label{fig:modelcomp}}
\end{figure}

The model-level statistics point the same way. Model~B explains 10.5\% of the
variance in training against 6.8\% for Model~A, and, more importantly, retains
far more of that under cross-validation: CV~$R^2 = 0.061$ for the local model
against $0.022$ for the global one, a factor of 2.8. The local model retains
about 60\% of its $R^2$ between training and held-out data, whereas the global
model retains about a third. Both the absolute performance and the generalization gap favor the local
measurement (Figure~\ref{fig:modelcomp}).

For context, the corresponding full model in Paper~I, fitted to all 5,105
clusters with the galaxy-wide sSFR, had CV~$R^2 = -0.017$, that is, worse than
predicting the sample mean for every cluster. The progression from
$-0.017$ to $0.022$ to $0.061$ across Figure~\ref{fig:modelcomp} should be read
with care. Only the second step is a controlled comparison: center and right
panels use the identical 325 clusters and differ solely in the environmental
variable. The first step changes three things at once, the age cut, the
restriction from 31 galaxies to four, and the sample size, so it cannot be
attributed to any one of them.

\begin{deluxetable*}{lcccccc}
\tablecaption{Regression Coefficients for the 325 Reliable Young Clusters
\label{tab:coeffs}}
\tablehead{
\colhead{} & \multicolumn{3}{c}{Model A: global sSFR} &
\multicolumn{3}{c}{Model B: local $\Sigma_{\rm SFR}$} \\
\cline{2-4} \cline{5-7}
\colhead{Parameter} & \colhead{$\beta$} & \colhead{$\sigma_\beta$} &
\colhead{$p$} & \colhead{$\beta$} & \colhead{$\sigma_\beta$} & \colhead{$p$}
}
\startdata
Intercept        & $-0.097$ & 0.543 & 0.86     & $0.076$ & 0.303 & 0.80    \\
$\log_{10} M$    & $0.123$  & 0.029 & $<0.001$ & $0.091$ & 0.028 & 0.0014  \\
$\log_{10} \tau$ & $0.078$  & 0.038 & 0.044    & $0.105$ & 0.038 & 0.0062  \\
Environment term & $0.042$  & 0.050 & 0.40     & $0.079$ & 0.021 & 0.00025 \\
\enddata
\tablecomments{Both models are fitted to the same 325 clusters and differ only
in the environmental term. Variance inflation factors are 1.07, 1.01 and 1.07
for Model~A and 1.07, 1.04 and 1.09 for Model~B, so multicollinearity is not
driving these results.}
\end{deluxetable*}

\begin{deluxetable}{lccccc}
\tablecaption{Model Performance on the 325 Reliable Young Clusters
\label{tab:performance}}
\tablehead{
\colhead{Environment term} & \colhead{$\beta_3$} & \colhead{$p$} &
\colhead{Training $R^2$} & \colhead{CV $R^2$} & \colhead{RMSE} \\
\colhead{} & \colhead{} & \colhead{} & \colhead{} & \colhead{} & \colhead{(dex)}
}
\startdata
Galaxy sSFR (global)              & 0.042 & 0.40    & 0.068 & 0.022 & 0.272 \\
$\Sigma_{\rm SFR}$ (local, $\Ha$) & 0.079 & 0.00025 & 0.105 & 0.061 & 0.266 \\
\enddata
\tablecomments{Both models also include $\log_{10} M$ and $\log_{10}\tau$.
CV $R^2$ is the mean coefficient of determination across five held-out folds.}
\end{deluxetable}

\subsection{How the Other Coefficients Shift} \label{sec:shifts}

Introducing the local environmental term does more than add a fourth
coefficient; it changes the two that were already there, and the direction of
those changes is informative.

The mass coefficient falls from $0.123$ in Model~A to $0.091$ in Model~B, a
decrease of 26\%. A plausible reading is that part of what mass appeared to
explain was in fact a proxy for environment. More massive clusters form
preferentially in denser, more actively star-forming regions, so in a model that
contains no local environmental term the mass coefficient absorbs some of that
environmental signal. Making the environment explicit returns it.

The age coefficient moves the other way, rising from $0.078$ to $0.105$, an
increase of 35\%. If clusters expand as they age, the age term should be
positive, and it becomes cleaner once environmental variation that was
previously confounded with age has been accounted for. Both shifts are
consistent with the local term capturing real structure in the data rather than
noise, though with 325 clusters neither shift is individually decisive.

\subsection{Within Galaxies or Between Them?} \label{sec:within}

With only four hosts, and 93\% of the sample drawn from two of them
(Table~\ref{tab:galaxies}), a pooled regression could in principle produce this
result without any cluster-scale physics. If NGC~1566 happened to have both a
higher mean $\SSFR$ and systematically larger clusters for some unrelated
reason, the pooled fit would show a positive $\SSFR$ coefficient that was purely
a difference between galaxies. Two tests separate these possibilities.

First, Model~B was refitted with galaxy fixed effects, that is with an indicator
variable for each host, which removes all between-galaxy variance and leaves
only variation among clusters within the same galaxy. The $\SSFR$ coefficient is
essentially unchanged: $\beta_3 = 0.068$ with $p = 0.0013$, against $0.079$ and
$p = 0.00025$ for the pooled fit.

Second, the model was fitted separately within each galaxy. Only NGC~628
($n = 176$) and NGC~1566 ($n = 126$) contain enough clusters to support an
independent fit; NGC~1433 ($n = 19$) and NGC~3351 ($n = 4$) do not, and are not
reported. Both fittable galaxies show the effect on their own, with the same
sign: $\beta_3 = 0.067$ ($p = 0.040$) for NGC~628 and $\beta_3 = 0.105$
($p = 0.0026$) for NGC~1566.

The environmental signal is therefore present among clusters within a single
galaxy and is not an artifact of differences between the four hosts. One further
point follows from Table~\ref{tab:galaxies}: these two galaxies lie at 8.8 and
15.6~Mpc, so the fixed $1\arcsec$ aperture averages $\SSFR$ over radii of
42.7~pc and 75.6~pc respectively. That the effect appears in both, at comparable
strength, argues that it is not tied to one particular physical aperture scale.

\subsection{Robustness to the Reliability Cut} \label{sec:robustness}

The 325-cluster sample is the conservative one, obtained by discarding 293 of
the 618 young clusters on the basis of the \citet{BrownGnedin2021} quality
flags. It is worth asking whether the result is an artifact of that cut.

It is not. On the full 618-cluster sample, which includes clusters with less
certain radius and mass measurements, the local $\SSFR$ signal is stronger, not
weaker: the coefficient rises to $0.146$ and CV~$R^2$ to $0.173$, against $0.057$
and $0.080$ for the global sSFR on the same 618 clusters
(Table~\ref{tab:robust}). The effect is therefore attenuated, not manufactured,
by the quality cut, and the ordering of local over global is unchanged. We do
not attempt to explain why the signal is stronger in the noisier sample; the
relevant point for this paper is the direction of the change. We quote the
325-cluster numbers throughout, so that no conclusion here rests on clusters
that \citet{BrownGnedin2021} would have excluded.

\begin{deluxetable}{lcccc}
\tablecaption{Local $\Sigma_{\rm SFR}$ Across Two Sample Definitions
\label{tab:robust}}
\tablehead{
\colhead{Sample} & \colhead{$N$} & \colhead{Environment} &
\colhead{$\beta_3$} & \colhead{CV $R^2$}
}
\startdata
All young           & 618 & Global sSFR                & 0.057 & 0.080 \\
All young           & 618 & Local $\Sigma_{\rm SFR}$   & 0.146 & 0.173 \\
Reliable only       & 325 & Global sSFR                & 0.042 & 0.022 \\
Reliable only       & 325 & Local $\Sigma_{\rm SFR}$   & 0.079 & 0.061 \\
\enddata
\tablecomments{In both samples the local measurement outperforms the global one
by a wide margin. The 325-cluster rows are adopted for the primary results.}
\end{deluxetable}

\section{Discussion} \label{sec:discussion}

\subsection{Why the Scale of Measurement Matters} \label{sec:scale}

The central result of this paper is that two variables intended to measure the
same physical thing, the star formation environment of a cluster, behave
completely differently depending on the scale at which they are measured. On the
same 325 clusters, in the same regression, the galaxy-wide version is
indistinguishable from zero and the local version is significant at
$p = 2\times10^{-4}$.

This is what one would expect if the relevant physics is local. Clusters form
inside individual giant molecular clouds, and the gas surface density, pressure,
turbulent velocity, and feedback environment of those clouds vary by more than
an order of magnitude within a single galaxy \citep{Sun2018, Chevance2020}.
A galaxy-wide sSFR assigns the same value to a cluster forming in a quiescent
outer-disk cloud and to one forming in a bright \ion{H}{2} complex in a spiral
arm. Averaging over that variation does not merely weaken the environmental
signal; it removes it by construction, because within a galaxy the variable is
constant. The null result of Paper~I is therefore best read not as evidence
against an environmental effect but as evidence that a galaxy-averaged variable
cannot detect one.

That an environmental variable matters at all is the qualitative expectation of
models in which the natal environment sets the initial structure of a cluster
\citep{ChoksiKruijssen2021}. The sign, however, is opposite to the specific
prediction of those models: we find \emph{larger} radii in regions of higher
$\SSFR$, whereas $r_h \propto \Sigma_{\rm gas}^{-1/2}$ implies \emph{more
compact} clusters in denser gas. This apparent conflict should not be
overinterpreted, since $\SSFR$ is not $\Sigma_{\rm gas}$ and the
relation between them is set by the star formation law, and since our clusters
have already undergone up to 10~Myr of gas expulsion and early dynamical
evolution, which is expected to be more violent in regions of intense feedback
\citep{Krumholz2019}. Distinguishing these possibilities requires the molecular
gas surface density directly, which is discussed below.

\subsection{Mass Remains Dominant} \label{sec:massdominant}

The environmental effect reported here is secondary, not primary. The mass
coefficient in Model~B is $0.091$ against $0.079$ for the environment, and the
total explanatory power of even the best model is modest: CV~$R^2 = 0.061$ means
that roughly 6\% of the out-of-sample variance in cluster radius is accounted
for. The radius-mass relation retains an intrinsic scatter of about 0.25~dex
\citep{BrownGnedin2021}, arising from cluster-to-cluster variation in formation
conditions, dynamical history, and measurement uncertainty, and no model built
from mass, age, and a single environmental scalar will approach explaining it.
The claim here is narrow and specific: the local birth environment leaves a
detectable, reproducible imprint on cluster structure, over and above mass.

\subsection{Cross-Validation as a Guard Against Overfitting}
\label{sec:cv}

A methodological point carries over from Paper~I and is reinforced here. In
Paper~I, the full multivariate model was favored by both the Akaike and Bayesian
information criteria and returned a significant partial $F$-test, yet its
CV~$R^2$ was negative. In-sample statistics are easy to satisfy with a large
sample, and they did not detect that the additional predictors were fitting
noise. Only held-out evaluation did.

The present result passes that same test. Model~B is significant \emph{and}
generalizes, retaining about 60\% of its $R^2$ out of sample against about a
third for Model~A. We
would encourage the reporting of cross-validated statistics alongside
significance tests in scaling-relation work generally, since the two can point
in opposite directions.

\subsection{Limitations} \label{sec:limitations}

Several limitations bound the strength of these conclusions.

\emph{Sample size and galaxy coverage.} The analysis rests on 325 clusters, and
although four galaxies contribute, NGC~628 and NGC~1566 supply 93\% of them.
Section~\ref{sec:within} shows that the environmental effect is present within
each of those two galaxies separately and survives galaxy fixed effects, so it
is not a between-galaxy artifact; but two galaxies remain a narrow and not
obviously representative slice of the local population. Nothing here establishes
that the relation holds in galaxies of different mass, metallicity, or dynamical
state, and the modest CV~$R^2$ values reflect this limited sample as well as the
large intrinsic scatter of the underlying relation.

\emph{Dust extinction.} $\Ha$ fluxes were not corrected for internal extinction.
Dust attenuation correlates with gas surface density, so uncorrected $\Ha$
systematically underestimates $\SSFR$ most severely in exactly the dense regions
of greatest interest. This should flatten the fitted relation rather than create
one, making our coefficient a conservative estimate, but the correction should
be applied before the coefficient is interpreted quantitatively. Balmer
decrements are available in the PHANGS--MUSE data products and would permit a
\citet{Calzetti2000} correction.

\emph{Fixed angular aperture.} A $1\arcsec$ aperture subtends different physical
radii in galaxies at different distances (Table~\ref{tab:galaxies}), so $\SSFR$
is averaged over different physical scales from galaxy to galaxy. Repeating the
measurement with a fixed physical aperture would test whether this matters.

\emph{Timescale of the tracer.} $\Ha$ traces only the last ${\sim}10$~Myr, which
is what makes it appropriate for this sample but also what confines the analysis
to it. Extending the test to older clusters would require a tracer of the birth
environment that survives cluster migration, and would reintroduce the problem
that a cluster's present position need not reflect where it formed.

\emph{$\Ha$ as a proxy for natal gas.} $\Ha$ measures ionized gas produced by
stars that have already formed, not the molecular gas from which the cluster
formed. Ionizing photon leakage, and the clearing of gas by feedback within a few
Myr \citep{Chevance2020}, both add scatter. Repeating this analysis with CO-derived
molecular gas surface densities from PHANGS--ALMA \citep{Leroy2021} would test the
\citet{ChoksiKruijssen2021} prediction in the variable in which it was made.

\subsection{Future Directions} \label{sec:future}

The most direct extension is to replace $\Ha$ with $\Sigma_{\rm mol}$ from
PHANGS--ALMA \citep{Leroy2021}, which both tests the theoretical prediction in
its native variable and removes the extinction and leakage concerns above. A
second is to enlarge the overlap sample: any expansion of integral field
coverage of LEGUS galaxies would improve the statistical power, which is
currently the binding constraint. A third is to add local metallicity, also
available from PHANGS--MUSE, as a second local environmental variable. Finally,
testing whether the effect strengthens in extreme environments such as
starbursts and interacting systems would probe the relation over a much wider
dynamic range in $\SSFR$ than the four star-forming disks studied here
provide.

\section{Summary} \label{sec:summary}

We tested whether the star formation environment predicts young star cluster
radius when it is measured locally rather than as a galaxy-wide average, using
$\Ha$ maps from PHANGS--MUSE matched to the LEGUS cluster catalog of
\citet{BrownGnedin2021} in the four galaxies common to both surveys. Our
findings are:

\begin{enumerate}

\item For 325 clusters younger than 10~Myr with reliable radius and mass
measurements, the local star formation rate surface density is a significant
predictor of effective radius ($\beta = 0.079$, $p = 2\times10^{-4}$) in a model
containing mass and age.

\item The galaxy-wide specific star formation rate is not a significant
predictor for the same 325 clusters ($\beta = 0.042$, $p = 0.40$), consistent
with the null result of Paper~I on the parent sample. The local coefficient is
1.9 times the global one.

\item The local model generalizes 2.8 times better under 5-fold cross-validation
(CV~$R^2 = 0.061$ versus $0.022$), retaining about 60\% of its explanatory power
out of sample against about a third for the global model.

\item The effect is not a between-galaxy artifact. With galaxy fixed effects the
coefficient is $0.068$ ($p = 0.0013$), and fitted separately the two galaxies
with sufficient clusters both show it: NGC~628 gives $0.067$ ($p = 0.040$) and
NGC~1566 gives $0.105$ ($p = 0.0026$). These two galaxies also lie at different
distances, so the $1\arcsec$ aperture samples 42.7 and 75.6~pc respectively,
indicating the result is not tied to one physical aperture scale.

\item Including local $\SSFR$ reduces the mass coefficient by 26\% and raises
the age coefficient by 35\%, suggesting that some of the apparent mass
dependence of cluster radius encodes the environment in which massive clusters
preferentially form.

\item The effect is not an artifact of the quality cut: on the full 618-cluster
young sample the local coefficient rises to $0.146$ with CV~$R^2 = 0.173$.

\end{enumerate}

Taken with Paper~I, these results support a picture in which mass, through virial
equilibrium, sets the primary radius-mass scaling, while the local birth
environment leaves a secondary but detectable imprint. A galaxy-wide average
cannot recover that imprint, because within a galaxy it does not vary. Where a
cluster forms within its host galaxy matters, but only if it is measured where
the cluster formed.

\begin{acknowledgments}
I thank Gillen Brown for mentorship, guidance, and constructive feedback
throughout this research, and for making the cluster catalog of
\citet{BrownGnedin2021} publicly available.

This work is based on observations made with the NASA/ESA \textit{Hubble Space
Telescope}, obtained from the Data Archive at the Space Telescope Science
Institute, which is operated by the Association of Universities for Research in
Astronomy, Inc., under NASA contract NAS 5-26555. These observations are
associated with program \#13364 (LEGUS).

Based on observations collected at the European Southern Observatory under ESO
programmes 1100.B-0651, 095.C-0473, and 094.C-0623 (PHANGS-MUSE; PI:
Schinnerer), as well as 094.B-0321 (MAGNUM; PI: Marconi), 099.B-0242,
0100.B-0116, 098.B-0551 (MAD; PI: Carollo) and 097.B-0640 (TIMER; PI: Gadotti).
The PHANGS--MUSE data products used here were retrieved through the CANFAR
science portal.
\end{acknowledgments}

\facilities{HST, VLT:Yepun (MUSE)}

\software{
  astropy \citep{Astropy2022},
  photutils \citep{photutils},
  matplotlib \citep{matplotlib},
  numpy \citep{numpy},
  scikit-learn \citep{sklearn},
  scipy \citep{scipy},
  statsmodels \citep{statsmodels}
}

\section*{Data Availability}

The LEGUS cluster catalog of \citet{BrownGnedin2021} is publicly available at
\url{https://github.com/gillenbrown/LEGUS-sizes}. PHANGS--MUSE data products are
public and were obtained from the CANFAR science portal. All analysis code used
in this work, including the $\Ha$ photometry, the conversion to $\SSFR$, and the
regression and cross-validation pipeline, is available at
\url{https://github.com/michelle-sebastian/starcluster}.

\bibliography{refs}
\bibliographystyle{aasjournalv7}

\end{document}